# HIGHLY STABLE COMMON-PATH QUANTITATIVE PHASE MICROSCOPE FOR BIOMEDICAL IMAGING

AZEEM AHMAD[1,2,3], VISHESH DUBEY[1,2], ANKIT BUTOLA[1], BALPREET SINGH AHLUWALIA[2], DALIP SINGH MEHTA[1,4]

[1]*Department of Physics, Indian Institute of Technology Delhi, Hauz Khas, New Delhi 110016, India*
[2]*Department of Physics and Technology, UiT The Arctic University of Norway, Tromsø 9037, Norway*
[3]*ahmadazeem870@gmail.com*
[4]*mehtads@physics.iitd.ac.in*

**Abstract:** High temporal stability is the primary requirement of any quantitative phase microscope (QPM) systems for the early stage detection of various human related diseases. The high temporal stability of the system provides accurate measurement of membrane fluctuations of the biological cells, which can be good indicator of various diseases. We developed a single element highly stable common-path QPM system to obtain temporally stable holograms of the biological specimens. With the proposed system, the temporal stability is obtained ~ 15 mrad without using any vibration isolation table. The capability of the proposed system is demonstrated on USAF resolution chart, polystyrene spheres (dia. 4.5 μm) and human red blood cells (RBCs). The membrane fluctuation of healthy human RBCs is further successfully measured and found to be equal to 63 nm. Contrary to its counterparts, present system offers energy efficient, cost effective and simple way of generating object and reference beam for the development of common-path QPM.

## 1. Introduction

Quantitative phase microscopy (QPM) is a powerful technique, which works on the principle of interferometry [1]. QPM has the capability to measure the phase maps and subsequently the refractive index and dry mass density of the biological cells [1-3]. To produce interference, there must be at least two coherent beams, one reference and one object beam. In the past, several attempts have been made by many researchers to produce two coherent beams. In most of the QPM systems, i.e., Michelson [4], Linnik [5-7], Mach-Zehnder [8], and Mirau interferometer geometries [2, 9, 10] are generally employed to produce two coherent beams, one from the reference arm and other from the object arm. These optical configurations for QPM, being non-common path in nature, suffer from several serious problems: time-varying phase noise due to vibration, temperature gradient, and air flow which deteriorate the stability of QPM [3, 11]. These issues limit the application of QPM systems for the study of live cells dynamics, i.e., measurement of membrane fluctuations, which is a good indicator for the detection of several diseases [11].

During the last decade, various common-path QPM techniques have been developed to minimize the temporal phase noise [1, 3, 11-16]. Diffraction phase microscopy (DPM) [3, 12] or spatial light interference microscopy (SLIM) [13] employ diffraction grating and SLM, which makes the system costly. The diffraction grating diffracts the input beam into several orders, which reduces the diffracted beam intensity into zero and +1 order by several orders of magnitude. The diffracted intensity into these two orders can be enhanced by employing blaze grating, which diffracts most of the incident light into either the zero or positive first order at the cost of 4% reflection from the grating surface and very little diffracted light into the negative first and higher orders [11]. However, diffraction grating limits the use of the incident light wavelength. It is optimized only for a particular wavelength for which it diffracts the maximum optical power into a specified diffraction order. To date, it utilizes the maximum incident optical power at the cost of aforementioned constraints, to generate highly stable holograms or interferograms. Moreover, the grating pitch decides the parameters (magnification and numerical aperture) of the objective lens to be used. To make the technique fully functional, approximately 3 – 4 grating lines should pass through the airy disk produced by the objective lens [11]. The diffraction grating is thus in-directly coupled to the chosen objective lens of given magnification and resolution. Therefore, if the objective lens is changed with different magnification or numerical aperture (NA), the diffraction grating must be changed accordingly to achieve optimal performance of the system [11]. This increases the complexity, i.e., the need of re-alignment and the practicality of the system.

In 2012, lateral shearing digital holographic microscopy (DHM) technique is also developed to achieve high temporal stability [14, 16]. However, this technique also has its own limitation, e.g., amount of shear between the reflected wave fronts generated due to the front and back surface of the shear plate has to be greater than the size of the object under study. Therefore, it cannot be implemented for closely spaced biological or industrial objects, i.e., sample has to be spatially dispersed. Further, the technique presented in Ref. [14] utilizes only 4 – 8% intensity of the input beam. To overcome these challenges a simple, comparatively energy efficient and cost-effective method is required.

In the present work, a highly stable QPM system is developed to overcome aforementioned issues. Two nearly common-path coherent beams are generated with the combination of 6 mm thick glass plate and pinhole assembly, thus, provides high temporal stability to the QPM system. The glass plate has ~ 20 % front and ~ 100 % back surface coating. The back surface reflected beam which carries high intensity is passed through 30 µm pinhole to generate the reference beam from the object beam. The temporal stability of the proposed system is measured to be equal to ~ 15 mrad without using vibration isolation table. The experiments are conducted on USAF resolution target, polystyrene sphere (dia. 4.5 µm) and human red blood cells (RBCs). The membrane fluctuation of healthy RBCs is further quantified and found to be greater than the system's temporal stability (~ 15 mrad.). The standard deviation of human RBC's membrane fluctuation is measured to be equal to 63 nm. The newly developed phase microscope can be easily integrated with dynamical behavior studying techniques [5, 17-20].

## 2. Materials and methods

### 2.1 Highly stable common-path quantitative phase microscope

A highly stable, comparatively energy efficient and cost effective common–path quantitative phase microscope is illustrated in Fig. 1. In the present technique, a novel way for generating reference beam and the object beam of approximately equal intensity is presented. This is achieved by using a combination of back surface silver coated optical glass plate and a pinhole assembly. A highly coherent light beam generated by He- Ne laser source (power ~ 15 mW and coherence length ~ 15 cm) is passed through a spatial filtering unit to generate a clean beam. The spatially filtered diverging beam is further made collimated by employing a collimating lens L1 into the beam path. Thus, the sample is illuminated from a collimated light beam, which avoids inaccuracy in phase measurement of the specimen. The transmitted light through the specimen is finally collected by 50X (0.7 NA) microscope objective (MO) lens, which is further imaged at the image plane (IP) using a tube lens L2.

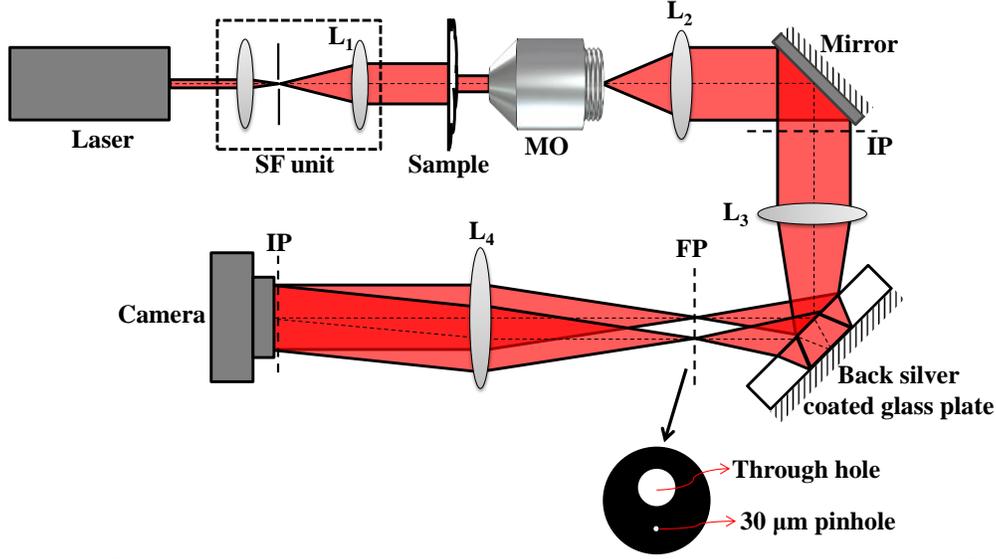

Fig. 1. (a) The schematic depiction of the transmission-mode common-path digital holographic quantitative phase microscope for biomedical imaging. SF unit: Spatial filtering unit; MO: Microscope objective; IP: Image plane. The amplitude mask having through-hole (dia. 6 mm) and pinhole (dia. 30 μm) situated at Fourier transform plane.

The object beam coming from the microscopic imaging system is then converged at Fourier plane (FP) by inserting a condenser lens L3. The two beams are generated utilizing 20% front and 100% back surface silver coated optical glass plate (Fig. 1), which produces two beams (reflection from the front and back surface of the glass plate) having high energy in one of the beam (reflected from the back coated surface). These two beams are further passed through a special kind of metal disk having two through holes. In one of the hole, a pinhole of 30 μm diameter is inserted. This pinhole containing disk is placed into the beam path in a way such that high energy beam should pass through the pinhole, which blocks most of the intensity (approximately 70 %) of the beam, thus generating a reference beam of approximately same intensity as that of object beam (20% reflection from the front surface). These two beams having almost equal intensity re-combine at the detector plane to generate high contrast spatially modulated interferograms of the specimen under study. The spatially modulated interferograms are further processed to produce specimen phase images, which contain information about refractive index and height about the sample.

## 3. Results and discussion

### 3.1 Temporal phase stability

The spatially modulated interferograms, generated due to the superposition of spatially filtered beam 'II' (after passing through pinhole) and unfiltered beam 'I', are further processed through the Fourier transform based phase retrieval algorithm [21] to measure the phase shift introduced by the specimen. The measured phase shift is related to the refractive index and height of the sample as:

$$\Delta\varphi(x, y) = \frac{2\pi}{\lambda} \Delta n\ h(x, y) \qquad (1)$$

Where $\lambda$ is the illumination wavelength, $\Delta n$ is the refractive index difference between the sample and surrounding medium and h(x,y) is sample thickness.

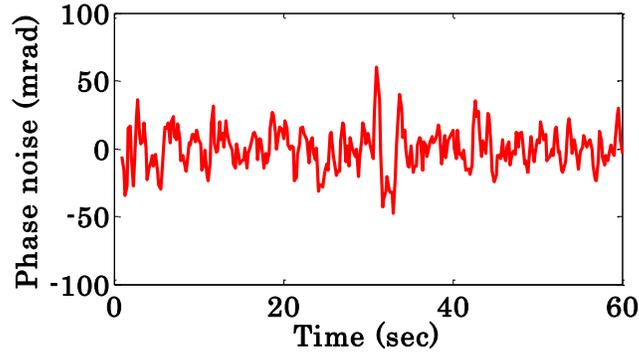

Fig. 2. Temporal phase noise variation plot of the developed phase microscope under ambient environmental fluctuation, varying over a range of [– 15, 15] mrad.

In order to illustrate the temporal stability of the present system, a series of time lapsed interferograms having size 1390 × 1040 pixels without any sample were recorded (Visualization 1). The time lapsed interferograms were further processed to measure the phase variation at a single spatial location of the retrieved phase map with respect to time. The temporal phase noise distribution of the present system is then calculated as

$$\Delta\phi(x, y, t) = \phi(x, y, t) - \phi(x, y, t = 0) \qquad (2)$$

Figure 2 depicts the phase variation of the developed phase microscope, over a range of [– 15, 15] mrad, under the ambient environmental fluctuations.

### 3.2 USAF resolution test target

The developed DHM system is first tested by performing experiments on USAF resolution test target. The resolution chart is placed into the beam path as shown in Fig. 1. The proposed system generates a high fringe density hologram of the USAF chart as illustrated in Fig. 3a. Figure 3b depicts the enlarged view of the region marked with red dotted box shown in Fig. 3a. The Fourier transform based reconstruction algorithm [5, 21] is implemented to recover amplitude information as presented in Fig. 3c.

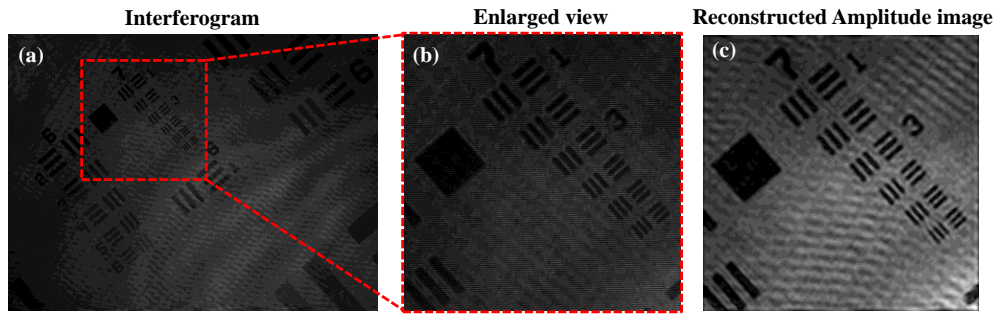

Fig. 3. (a) Fourier hologram of USAF resolution chart captured using the present common-path QPM. (b) Enlarged view of the small portion of the interferogram of Fig. 3a enclosed by red dotted box. (c) Reconstructed amplitude image of resolution target obtained using Fourier transform method.

### 3.2.1. Quantitative phase imaging

#### 3.2.1.1. QPI of polystyrene spheres

To test the phase measurement accuracy of the present system, experiment is done on the polystyrene spheres (Polysciences, Inc. Catalog # 17135) with diameter of 4.5 ± 0.152 μm and refractive index of 1.59 at 632 nm wavelength. The polystyrene spheres are first washed 2× with distilled water and then immersed in ethanol. A small volume of polystyrene solution pipetted into a 3×3 mm2 polydimethylsiloxane (PDMS) chamber attached with microscopic glass slide and left for 5 - 10 minutes for the evaporation of ethanol. Then a drop of immersion oil (RI ~ 1.51) is put onto the dried polystyrene sphere's monolayer to avoid the lensing effect caused by its spherical nature.

The sample is sealed from the top with a 170 μm thick glass cover slip. The sample is placed under the developed QPM for phase measurement and subsequently height measurement of polystyrene spheres. Figure 4a illustrates full-frame 2D interferogram of polystyrene spheres generated by the present setup. The magnified view of the region marked with red dotted box is depicted in Fig. 4b and subsequently utilized to measure phase/height map of polystyrene spheres. The 2D and 3D views of polystyrene sphere's phase map are presented in Figs. 4c and 4d, respectively.

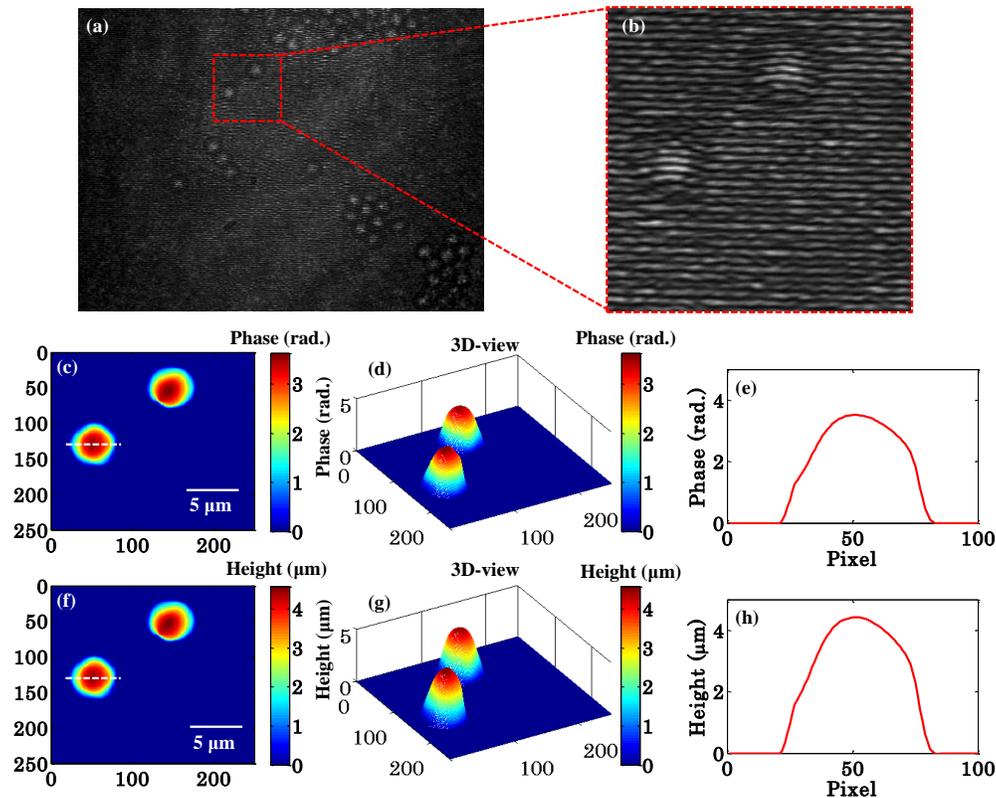

Fig. 4. (a) Full-frame interferometric image of polystyrene spheres (dia. ~ 4.5 ± 0.152 μm) captured using the present common-path QPM. (b) Enlarged view of the small portion of polystyrene spheres interferogram of Fig. 4a enclosed by red dotted square. (c-e) 2D-view, 3D-view and corresponding line profile (along white dotted line) of reconstructed phase image of polystyrene spheres obtained using Fourier transform method. (f-h) 2D-view, 3D-view and corresponding line profile (along white dotted line) of the height map of polystyrene spheres obtained from phase map. Color bar shows the phase in rad and height in μm.

The line profile is then plotted along the white dotted line depicted in Fig. 4b is illustrated in Fig. 4e. The maximum phase value of the polystyrene sphere is found to be equal to 3.51 rad. Figures 4f and 4g exhibit the 2D and 3D views of the measured height map of the polystyrene spheres. The refractive indices of polystyrene spheres and outside medium (immersion oil) are considered to be equal to 1.59 and 1.51 at 632 nm wavelength for the calculation. The height profile of the polystyrene sphere along the white dotted line shown in Fig. 4f is depicted in Fig. 4h. The height of the polystyrene spheres is measured to be equal to 4.42 μm, which is found to be in a good agreement with the values (~ 4.5 μm) provided by the manufacturer.

*3.2.1.2. QPI of human RBCs*

To show the feasibility of the present system, human red blood cells (RBCs) were imaged. For the preparation of biological sample, fresh blood sample is collected from a healthy donor with skin puncture. The fingertip is, first, cleaned with 70% isopropyl alcohol before the skin puncture. The skin is then punctured with one quick stroke to achieve a good flow of blood from the fingertip. The first blood drop is wiped away to avoid the excessive tissue fluid or debris. The surrounding tissues are gently pressed until another blood drop appears. The blood drop is then put onto microscopic glass slide and spread using a cover slip to form a thin blood smear for interferometric recording.

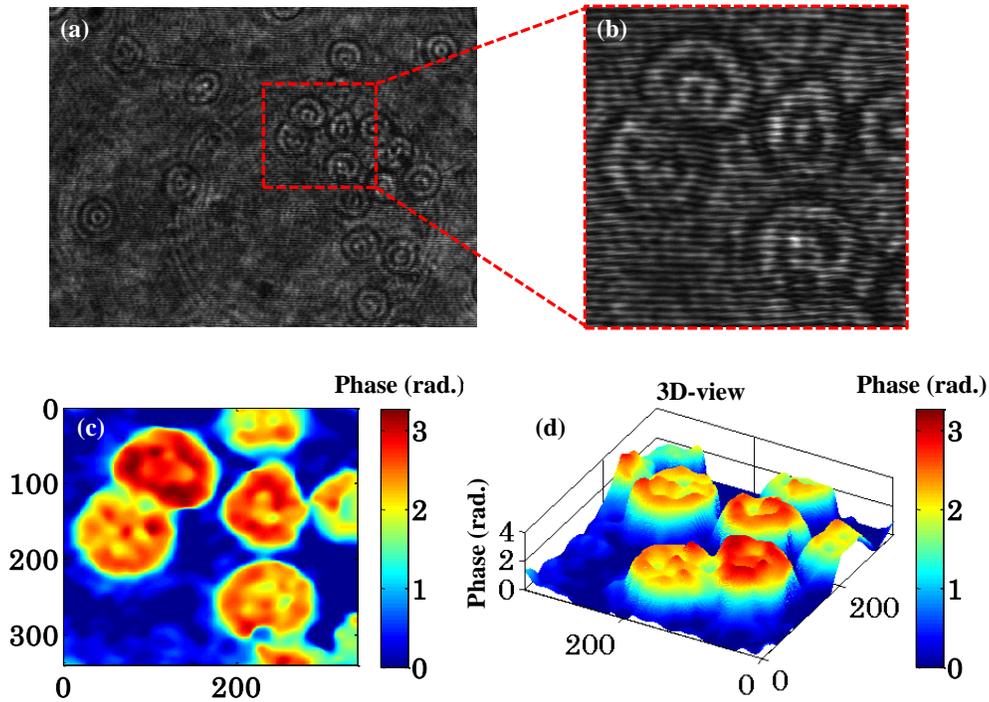

Fig. 5. (a) Full-frame interferometric image of human RBCs captured using the present common-path QPM. (b) Enlarged view of the small portion of RBCs interferogram of Fig. 5a enclosed by red dotted square. (c) 2D-view of reconstructed phase image of human RBCs obtained using Fourier transform method. (d) Pseudo 3D view of the phase map of human RBCs. Color bar shows the phase in rad.

Figure 5a depicts the interferogram of human RBCs captured by employing the developed phase microscope. To avoid the computational load, a small portion of the interferogram was selected shown in Fig. 5b and processed further using the Fourier transform based phase retrieval algorithm. Figure 5c and 5d represent the 2D and 3D view of the recovered phase map corresponding to the interferogram. The background phase from the recovered phase map was subtracted numerically by employing a modified polynomial fitting method [22]. The recovered phase map can be further utilized to measure several parameters related to cells such as cell's height, refractive index, cell's dry mass, hemoglobin content, hemoglobin concentration and membrane fluctuation etc.

### 3.2.2. Membrane fluctuation of human RBCs

To exhibit the potential application of high temporal stability of the proposed QPM system, membrane fluctuation of the healthy RBC is measured. Fresh blood sample is collected from the hospital in the Ethylenediaminetetraacetic acid (EDTA) containing tubes, which avoids the coagulation of RBCs. The blood sample was then diluted using phosphate buffer saline (PBS) and washed 2X to isolate the human RBCs from the rest of the blood components such as white blood cells, platelets and plasma. Further, sample was prepared on a microscopic glass slide and placed under the microscope for the measurement of RBC's membrane fluctuation.

To assess the RBC's membrane fluctuation measurement capability of the proposed system, we acquired a 1 min time-lapsed interferometric movie of human RBCs under ambient environmental fluctuations and without vibration isolation table (see Visualization 2). Each interferometric frame of the movie is processed by employing Fourier transform based phase recovery algorithm for the measurement of height fluctuations of RBCs. The refractive index of human RBC and outside medium (PBS) are considered to be equal to 1.42 and 1.33, respectively, during the calculation [23]. The average height map of human RBC corresponding to the time-lapsed interferometric movie is illustrated in Fig. 6a. The temporal variation of the RBC's height map can be seen in Visualization 3. For the measurement of membrane fluctuation, average height map is subtracted from each recovered instantaneous height map of RBC corresponding to each frame of the interferometric movie (Visualization 4).

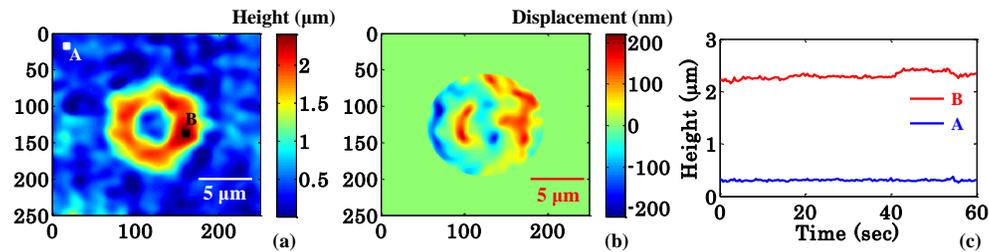

Fig .6. Membrane fluctuation measurement of healthy RBC. (a) Recovered average height map of human RBC corresponding to the time-lapsed interferometric movie (Visualization 2 and 3). (b) Respective instantaneous displacement map (color bar in nm) (see Visualization 4). (c) Height fluctuations of A, B during 60 seconds.

The instantaneous displacement map of the RBC is depicted in Fig. 6b, which is found to be in a close agreement with the previous results [24]. The height changes at two different positions are monitored over 60 s (Fig. 6c). The point on the glass slide (A) exhibits the temporal stability of the QPM system. The standard deviation (SD) of height fluctuation at point A is measured to be equal to 15.2 nm. The point chosen on the RBC membrane (B) presents the dynamic fluctuation in the RBC's membrane; the SD is measured to be equal to 63 nm. It is worth noting that the present system has the capability to quantify RBC's dynamic fluctuations under environmental fluctuations, which can be further employed for various disease detections.

## 4. Conclusion and discussion

The present method exhibits a novel way of generating reference beam and the object beam of approximately equal intensity from the sample beam passing through the microscope objective lens. This is achieved by using a combination of silver coated optical glass plate and a pinhole assembly. The optical configuration is unique and does not require neutral density filter to equalize the intensity of two beams, which makes it energy efficient. The phase noise of system under ambient environmental fluctuations found to be quite low (better than 15 mrad.) due to its common path nature. This can be further improved by utilizing vibration isolation table. In addition, there is no need of sparse, i.e., spatially dispersed sample due to the use of pinhole at the Fourier plane for the generation of reference beam.

The capability of the present system is exhibited by performing experiments on various industrial and biological specimens such as USAF resolution chart, polystyrene spheres and human RBCs. The membrane fluctuations of the healthy RBC is successfully quantified with the proposed system, which can be further employed for the detection of various diseases such as malaria, dengue and sickle cell anemia etc. Due to its common-path nature, the system can also be implemented with other optical functions like optical tweezers, waveguide trapping, microfluidics etc.

Multi-wavelength lasers can be combined into the beam path to perform simultaneous multi-wavelength study of the biological specimens. Moreover, the present invention is independent on magnification and NA of the objective lens used to image the specimen unlike DPM. This will find applications in acquiring phase images over large FOV first by using low magnification objective lens and then acquiring zoomed image of the point of interest using high magnification objective lens without the need of changing the grating in the light path.

**Funding:** The authors are thankful to Department of Atomic Energy (DAE), Board of Research in Nuclear Sciences (BRNS) for financial grant no. 34/14/07/BRNS. B.S.A acknowledges the funding from the European Research Council, (project number 336716) and Norwegian Centre for International Cooperation in Education, SIU-Norway (Project number INCP- 2014/10024). D.S.M acknowledges University Grant Commission (UGC) India for joint funding.